\documentclass[journal,compsoc,10pt]{IEEEtran}
\usepackage[utf8]{inputenc}
\usepackage[T1]{fontenc}    %
\usepackage{hyperref}       %
\usepackage{url}%
\usepackage{array,colortbl,multirow,multicol,ctable}

\usepackage{booktabs}       %
\usepackage{amsfonts}       %
\usepackage{nicefrac}       %
\usepackage{microtype}      %
\usepackage[ruled]{algorithm2e}
\usepackage[backend=biber,style=ieee]{biblatex}
\usepackage{amsmath,amsthm,amssymb}
\usepackage{mathtools} %
\usepackage{bm} %

\usepackage{cleveref}
\usepackage{caption}
\usepackage{subcaption}
\graphicspath{{./imgs/}}
\DeclareMathOperator{\MLP}{MLP}

\DeclareMathOperator{\AGG}{AGG}

\begin{document}
\title{Topology Aware Deep Learning for \\ Wireless Network Optimization}
\author{
  Shuai~Zhang,~\IEEEmembership{Student Member,~IEEE,}
  Bo~Yin,~\IEEEmembership{Student Member,~IEEE}
  Yu~Cheng,~\IEEEmembership{Senior Member,~IEEE}
  \IEEEcompsocitemizethanks{
    \IEEEcompsocthanksitem S. Zhang, B. Yin and Y. Cheng are with the Department of Electrical and Computer Engineering, Illinois Institute of Technology, Chicago, IL, 60616.\protect\\
    E-mail: \{szhang104, byin\}@hawk.iit.edu, cheng@iit.edu.
    }%
    }
\IEEEtitleabstractindextext{
\begin{abstract}
  Data-driven machine learning approaches have been proposed to facilitate wireless network optimization by learning latent knowledge from historical optimization instances. However, the existing study is limited to the situation with a static topology and extension to dynamically varying topologies is challenging. The fundamental difficulty is that there is no canonical representation for graphical data and thus hard for a machine to conveniently learn how the variation in topology may affect the network flow optimization. To address this issue, we leverage the graphical neural network (GNN) techniques and propose a two-stage topology-aware machine learning framework, which trains a graph embedding unit and a link usage prediction module jointly to discover links that are likely to be used in optimal scheduling. Some important learning techniques are also developed or discussed to ensure learning efficiency in this type of problem.
  The proposed approach is evaluated on a canonical wireless network flow problem with diverse network structures and realistic deployment scenarios, and achieves close-to-optimum solution quality with significant computation time-saving.
\end{abstract}

\begin{IEEEkeywords}
  Deep learning, wireless network optimization, topology representation
\end{IEEEkeywords}}
\maketitle

\IEEEraisesectionheading{\section{Introduction}}

\IEEEPARstart{O}{ver} decades, mobile communications and  wireless networking have been evolving into indispensable activities in human's everyday life. The recent proliferation of the Internet-of-Things (IoT) and the dense deployment of new-generation wireless network access points or base stations have made the scale and complexity of wireless networking growing explosively. The study of wireless network optimization, although legacy, still plays a key role in modern wireless networks; with the pressure of large scale high complexity, and more dynamics, the need of efficient computation algorithms that can optimize network resource allocation in an adaptive and timely manner is more urgent. 

Most network optimization tasks follow the paradigm of mathematical programming: 
given the constraints of resource budget or exclusive usage at a single time, how performance metrics or utility can be improved through judicious allocation and scheduling. 
Specifically, with the broadcast nature of wireless networks, at any given time, only a subset of communication links can be activated concurrently to mitigate interference \cite{jain2005impact, ramachandran2006interference, badia2008general}. 
With such a combinatorial interference structure, it is not uncommon that most performance optimization in wireless networks are NP-hard \cite{goussevskaia2010efficiency}. 
To tackle this challenge, studies on wireless network optimization in the past years focus on the development of approximation algorithms \cite{chafekar2008approximation, misra2010constrained, gandhi2012approximation}, based on the mathematical model of the performance goal and the constraints.
One drawback of this traditional model-based optimization approach is that the computation \emph{experience} gained from solving historical problem instances is wasted: 
whenever changes to the networking setting occur, 
(for example, user traffic pattern can drift with time, the set of active users may change due to mobility, or network topology can shift due to the adjustment of resources), the optimization procedure needs to be rerun \cite{liu_deep_2018}, and there is no difference in the process even if the new situation is almost identical to the previous network condition.

Inspired by the recent success of machine learning (ML), data-driven approaches receive a lot of attention in the area of wireless network optimization. One major thread of exploiting the historical data for network management adopts the methodology of reinforcement learning (RL), in which the experience replay technique allows an agent to learn a reasonable control policy from its past interactions with the environment. Research attempts along this line have developed RL-based algorithms to address a large range of network optimization problems, including access control~\cite{wang2018deep, yu2019deep}, network scheduling~\cite{zhang2017energy, chinchali2018cellular}, and traffic engineering~\cite{xu2018experience, chen2018auto}. These studies typically focus on the optimization tasks for a certain network layer, e.g., physical layer, link layer or network layer, where the target system can be conveniently modeled by a Markov decision process (MDP). The application of ML in the cross-layer optimization in wireless networks is quite limited.

In this paper, we focus on the network flow optimization over a generic multihop wireless network. Such research issues had been one key research area in the recent two decades, and are expected to still play an important role in emerging areas such as vehicular ad hoc networks and aerial access networks. Note that wireless network flow optimization involves the joint solution of link-layer scheduling and network-layer routing problems and does not incorporate a clean MDP structure for applying reinforcement learning. The paradigm of supervised learning seems a promising direction that may benefit network flow optimization. Specifically, works in this thread utilize historical problem instances that are solved by conventional algorithms
as training data, from which the trained machine can learn a mapping function to predict or facilitate computing the solution of a new problem instance \cite{wijaya2016neural, tang2017removing, sun2018learning}.  However, it sounds overambitious to train a machine that can directly predict the solution for a wireless network problem, considering the high dimension, heterogeneity, and combinatorics in optimization decision variables.    

A very interesting approach is proposed in \cite{ liu_deep_2018}, which integrates deep learning (DL) with wireless network flow optimization to further mitigate the computation complexity, beyond the traditional methods of approximation algorithms. More precisely, the optimization solutions from historical problem instances are collected as data to conduct supervised learning to train a DL model. After training, the machine will gain the capability 
to predict the usage likelihood of each link, given a new problem instance; links with low predicted values are then pruned off from the network, so that the effective optimization problem scale is reduced with only minor solution quality degradation.

We noticed that the current study in \cite{liu_deep_2018} is limited to the situation with a static topology since the network topology information is not processed with a proper model.
In this paper, we aim to take up the challenge of extending the methodology to a more generic scenario with dynamic topologies. Effective re-optimization over various topologies will be particularly useful if a wireless network needs to timely handle the dynamics due to user mobility, traffic pattern change, or adjustment of network resources.  A straightforward idea to incorporate topology into learning is feeding the topology information, in the form of the adjacency matrix, directly to the machine. However, the topology representation based on the adjacency matrix will be dependent on the specific node indexes.
With dynamic networks, the communication nodes may acquire new indexes due to mobility or update of resource allocation and the situation of the same topologies but with different node indexes will not be rare. The fundamental difficulty is there is no canonical representation for graph data, and to tell if two graphs representations are structurally equivalent is shown to be computationally difficult \cite{babai2016graph}.
So far graphical neural networks (GNN) provide new ways to address the challenge of representing graphical problem data in a form that facilitates the training and inference.
They have been shown to be useful in solving some known hard graph problems including the traveling salesperson problem \cite{wu2020comprehensive}.

Leveraging on that, we propose a topology-aware deep learning (TADL) framework for wireless network flow optimization.
In TADL, graph embedding techniques are developed to incorporate structure-level topological information so that its output representations have network structure information built into it which is independent of node or link indexes. We show that by learning on such a topology-aware representation of the problem data, neural networks can accurately infer the links to be used (and thus prune non-critical links), leading to a good tradeoff between problem scale reduction and close to an optimal solution, robustly over various network topologies and different commodity flow deployment scenarios. 

In summary, this paper incorporates three-fold innovative contributions
\begin{itemize}
    \item We design the TADL framework that integrates graph embedding, attention mechanism, and some specially tailored implementation techniques. TADL, when appropriately trained with enough data, can achieve robust link usage predication over various network topologies and different commodity flow deployment scenarios without retraining.
    \item We develop a few important learning techniques which are essential to ensure good performance in topology-aware learning, including the proper design of loss function, sample selection with curriculum training, and feasibility guarantee with link pruning.
    \item We present extensive numerical results with some insights that are revealed for the first time, to the best of our knowledge, including the new dimension of complexity-performance tradeoff enabled by TADL, the impact of the training data volume on the learning performance in the context of wireless network optimization, and the applicability of the learned capability in new settings without retraining.

\end{itemize}

The remainder of this paper is organized as follows. Section~\ref{systemmodel} concisely describes
the wireless network optimization problem and associated conventional algorithm, which is to be studied with a machine learning approach in this paper. Section~\ref{secTADL} presents
the proposed TADL framework and related implementation details.
Numerical results and performance evaluations are presented in Section~\ref{secNum}. Section~\ref{related} reviews more related work. Section~\ref{summary} concludes the paper.

\section{System Model}
\label{systemmodel}

 In this paper, we consider a multi-commodity flow (MCF) problem in a single-radio single-channel (SRSC) wireless network. Note that a generic multi-radio multi-channel (MRMC) wireless network can be mapped as a virtual SRSC with the multidimensional tuple modeling technique developed in \cite{li2010multi}.

\subsection{Network Model}

The SRSC network is represented by a directed graph $\mathcal{G}(\mathcal{N}, \mathcal{E})$, where $\mathcal{N}$ and $\mathcal{E}$ denote the sets of nodes and links, respectively.  A communication link $e\in\mathcal{E}$ exists from node $u$ to node $v$, denoted by tuple $(u ,v)$ if node $v$ is within the \textit{communication range} of node $u$. Each link has a physical transmission capacity, say $c(u, v)$, specifying the peak data rate this link is able to support.

We consider the protocol interference model \cite{guptaCapacityWirelessNetworks2000a}. A link can transmit, or become activated, only when no other active transmitting node (different from the sending node of this link) is within the \textit{interference range} of the receiving node. There is also the radio constraint that links sharing the same node cannot be activated simultaneously. The conflict relations among all the links can be characterized by a conflict graph \cite{jainImpactInterferenceMultiHop2005}.
An independent set (IS) over the conflict graph indicates a set of links that can be scheduled for transmission simultaneously.  

In the network, there exits $K$ commodity flow demands, denoted as $\mathcal{C}$. A commodity flow $k$ can be represented by the tuple $(s_k, d_k)$, with $s_k$ being the source node with a positive net outward data flow and $t_k$ the sink node with a positive net inward flow. Let $f_k(u, v)$ denote the amount of traffic flow associated with commodity $k$ on link $(u, v)$. For commodity $k$, the achievable throughput $r_k$ is the net flow out of the source node as
\begin{equation}
r_k = \sum_{v \in \mathcal{N}^+_{s_k}} f_k(s_k, v) -  \sum_{v \in \mathcal{N}^-_{s_k}} f_k(v, s_k),
\end{equation}
where $\mathcal{N}^-_v$ (resp. $\mathcal{N}^+_v$) denotes the set of in-neighbors (resp. out-neighbors) of $v$.

\subsection{Problem Statement}

We here consider the network flow problem of maximizing the minimum commodity flow in the network. In a wireless setting, a network flow problem involves not only routing decisions but also scheduling decisions in which ISs are activated in a time-multiplexing manner.
Besides the optimization variables $f_k(u, v)$ that specify the routing decisions, let $\mathcal{M}$ be the set of all ISs and $\alpha_m$ denotes the fraction of time scheduled to IS $m$. Those decision variables need to satisfy the following constraints.
\begin{itemize}
	\item Link capacity constraints: the sum of all the flows over a link does not exceed its capacity across all the activated time periods, i.e.,
	\begin{equation}\label{constr:link}
	 \sum_{k=1}^K f_k(u, v) \le \sum_{m \in \mathcal{M}}\alpha_m p_m(u, v), \quad \forall (u, v) \in \mathcal{E},   
	\end{equation}
	where $p_m(u, v)$ is defined as the effective capacity in an IS $m$.
	Its value is $c(u, v)$ if link $(u, v)$ is activated in $m$ and $0$ otherwise.
	\item Flow conservation constraints: for any commodity flow $k$, the amount of flow entering an intermediate node equals to that exits the node, i.e., 
	\begin{equation}\label{constr:flow}
	\sum_{u \in \mathcal{N}^-_v}f_k(u, v) = \sum_{u \in \mathcal{N}^+_v}f_k(v, u), \quad \forall v \ne s_k, d_k; \forall k.    
	\end{equation}

    \item Scheduling time constraint: the time fraction assigned to all ISs must sum to $1$, as
    \begin{equation}\label{constr:time}
     \sum_{m \in \mathcal{M}}\alpha_m = 1.   
    \end{equation}
\end{itemize}

The MCF problem can be formulated as follows.
\begin{align}\label{op:formulation}
\underset{\{f_k(u, v)\}, \{a_m\}}{\text{Maximize}}&  \quad z \\
 \quad \text{s.t.} \quad &\text{constraints} (\ref{constr:link}), (\ref{constr:flow}), (\ref{constr:time}), \nonumber \\
&  z \leq r_k, \quad \forall k \nonumber \\
&  f_k(u, v) \ge 0, \quad \forall (u, v) \in \mathcal{E}, k \nonumber \\
&  \alpha_m \ge 0, \quad m \in \mathcal{M} \nonumber
\end{align}

The problem described above has the form of linear programming, because the objective and constraints are linear functions.
However, the size of $\mathcal{M}$ is exponentially large and cannot be easily enumerated;
even to obtain one set of non-interfering links is equivalent to finding a graph coloring of links.
Therefore the problem is essentially a mixed-integer linear programming (MINLP) type with an exponential number of variables.

To the best of our knowledge, the most efficient approximation with guaranteed bound analysis is the delayed column generation (DCG) method  \cite{bertsimas_introduction_1997} \cite{cheng2014systematic}.
Starting from an initial set of ISs, it solves a series of partial problems and uses the dual solution to generate new ISs to add in the problem.
Following this procedural column generation, a solution sufficiently close to the optimum solution to the original problem is obtained.
In this way, the algorithm memory usage is saved and complexity can be controlled as a trade-off with the objective.
The reader is referred to the work \cite{cheng2014systematic} for a more detailed account of this method, which we use as a teacher algorithm in the later sections.
The key intuition behind it is that the final optimum result only makes use of a very small subset out of all the ISs, and most of the other ISs are given a zero time-share.

\begin{figure*}[!hbt]
	\centering
	\includegraphics[height=5cm]{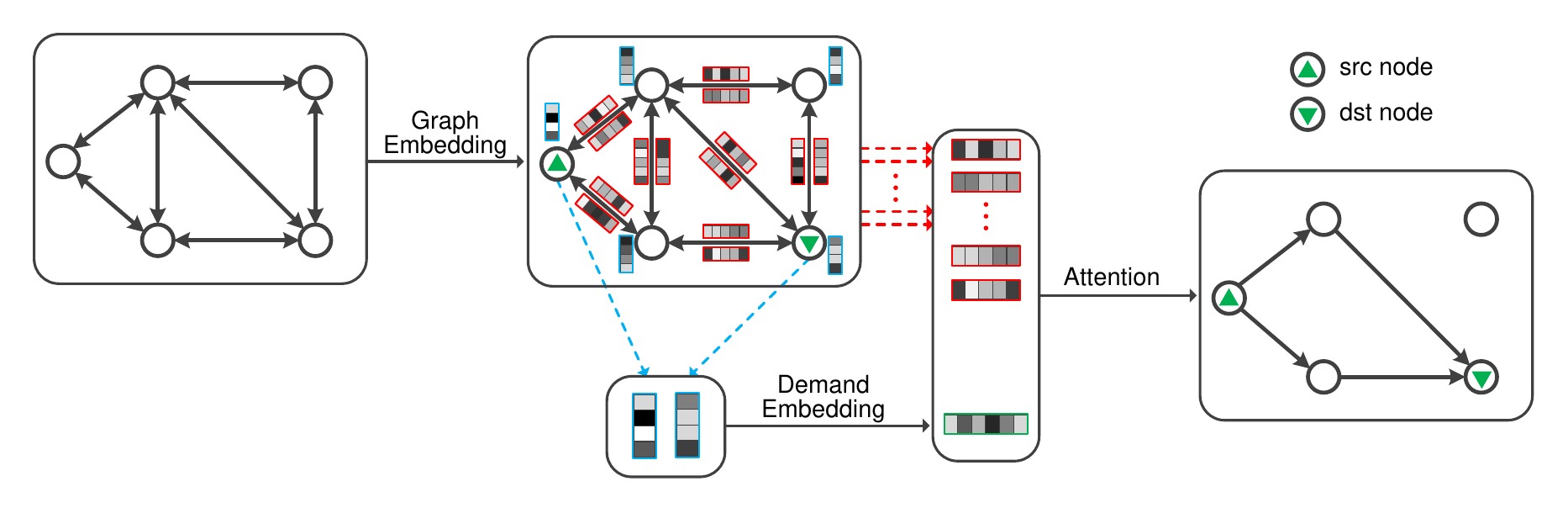}
	\caption{The proposed topology-aware deep learning (TADL) framework.}\label{fig:mlp-struc}
\end{figure*}

\section{Topology-Aware Deep Learning Framework \label{secTADL}}
Even when the optimization problem formulated in the previous section is solved with the efficient DCG method, the size of the IS space is often so large that it results in slow runtime when the number of links grows.
We adopt the basic idea proposed in \cite{liu_deep_2018} to leverage machine learning to mitigate the computation complexity.
Specifically, optimization solutions to historical problem instances will be used as training data to conduct supervised learning.
Then, given a new optimization instance, the trained model can predict and prune non-critical links from the original topology and thus feed a reduced-size formulation to the optimization solver.

We would like to emphasize that the brand new issue addressed in this section is to improve the basic training method proposed in \cite{liu_deep_2018} into a topology-aware framework, where a well-trained machine will robustly perform well over various network topologies. 
Specifically,
the proposed topology-aware deep learning (TADL) framework is illustrated in \cref{fig:mlp-struc}, which consists of a graph embedding unit, a demand embedding unit and an attention unit. 
The embedding units are mainly used to address the issues of topology representation as highlighted in the introduction. 
They learn a reasonable representation of a problem instance, including multiple link embedding vectors and a demand embedding vector. 
Such a representation is leveraged by the attention unit to identify network links that are likely to be used in an optimized way.

The remainder of the paper adopts some common mathematical notations.
We use a boldface uppercase/lowercase letter to represent a matrix/vector, respectively. 
We use $||$ to express vector concatenation and $\odot$ for element-wise multiplication.
$\text{LIN}$ representing a parameterized linear layer $\mathbf{y}(\mathbf{x}) = \mathbf{Ax} + \mathbf{b}$;
$\text{MLP}$ meaning a multi-layer perceptron, made up of several dense layers, each with individual non-linearity, and unless specified, uses batch norm at the last layer.
$\bm{D}$ and $\bm{A}$ are degree and adjacency matrix following the conventional graph theory definitions.
\newcommand{\yy}{\mathbf{y}}
\newcommand{\xx}{\mathbf{x}}

\subsection{Principles of Topology-aware embedding}

We consider the topology-aware processing to be an indispensable part of applying machine learning to network research,
because the lack of it leads to undesirable fitting on the network element ordering.
Assume that one directly learns from the usual representations like adjacency matrix, where an element $(i,j)$ is 1 if there is a connection between node $i$ and $j$ and 0 otherwise, it would be difficult to correctly represent the structure for two reasons.
One, the final decision is likely to depend on the specific order of the elements, as to learning algorithms, permuted adjacency matrices are different inputs.
If the permuted version does not appear in the training data, it is unlikely that they both correspond to the same output.
Second, assignment-matrix style representations are sparse for typical networks of non-trivially small size, since the number of usable links $|\mathcal{E}|$ is far below the number of possible node pairs $|\mathcal{N}|(|\mathcal{N}-1|)$.
Learning from sparse data is difficult to do with today's learning frameworks~\cite{Goodfellow-et-al-2016}.

The intrinsic difficulty is due to the fact that graphical structures lack inherent order:
unlike images or time series where there is a spatial or temporal ordering by which the representation is unique,
the nodes and edges information could be passed in any order to the learning algorithm while not changing the underlying mathematical object.
But neural networks are able to pick up \emph{any} pattern in the data, including those that arise from the particular ordering of the element.
While this serves well in other applications like image classification, it hurts the generalization ability in graph-based problems~\cite{zhang_interpreting_2017}.
On the other hand, generating a representation which corresponds to graph structure rather than its element ordering is reducible to the \textsc{Graph Isomorphism} problem in graph theory, which is non-trivial and still yet to be known to whether belong to the NP-class~\cite{schoning_graph_1988}.

Considering the importance and practical difficulty of differentiating graph structures, with our final goal --- predicting the link importance --- in mind, we  compromise by designing the embedding step to satisfy these properties instead:
\begin{enumerate}
  \item
    The output value should be constant with respect to the changes in the order of network nodes or links, if its dimension is not tied to the number of nodes or edges in the graph.
    This requires that we get the same evaluation of link importance regardless of how the node or edge changes its index.
  \item
    The output values should change its order in the same manner as the input if its dimension is tied to the number of nodes or edges in the graph.
    This is useful when we want to infer information on a per-node or per-link basis:
    it only makes sense that the prediction tracks the network elements it needs to represent.
  \item
    The vector representation of nodes or links should directly encode associated feature information.
    This item is necessary as in network applications, the associated information (e.g., node queue capacity, link strength etc) is as important as the graph structure itself in determining the output.
\end{enumerate}

The end result is a topology-based ``signature'' to each graph element, such that the learning algorithm could have access to not only the index information, but also the surrounding structure.

To achieve the goals stated above, we use the message-passing network~\cite{gilmer_neural_2017} approach.
It iteratively updates the edge and node embedding vectors based both on the feature vectors and the graph structure.
This process is also known in other works as the graph convolutional network~\cite{kipf2016semi}.
It bears a strong resemblance to how a network operates in the normal working state: each node sends to and receives from its neighbors and makes updates, and as a result the updated states is an implicit function of graph topology.
After a few rounds every node has partial information about the neighborhood it is in.

If this process is repeated on different scales, e.g., message passing from one neighborhood to another, then the local information gradually is refined into a high-level summary of the current global state.
The interconnection is encoded implicitly through multiple rounds of updates.
No matter how the nodes and links in the graph change their order in the graph, since topologically for each node or link, its neighbors remain unchanged.

Another consideration to ensure these properties is to adopt symmetry whenever possible.
When updating the embedding of network elements, care must be taken to ensure that each of them was processed in a symmetric manner.
Specifically, if the update is done with a parameterized model, it must remain the same across the whole set of similar elements.
Similarly, aggregation from multiple symmetrical elements, like nodes from the same graph, should treat them in the same manner.
This way the output will be decoupled from ordering and only related to network nodes' or links' feature and their interconnection.
Incidentally, Deep Convolutional networks (not to be confused with graph convolutional networks, though draws from similar high-level motivation, is different in both implementation and application) used in image learning embody such ideas.

\subsection{Encoding the topology and traffic demand}

The input to this model is a graphical representation of the problem
$(\mathcal{G}, \mathbf{V}, \mathbf{E})$, i.e., the graph itself and the set of node and link attributes.
Matrix 
$\bm{V} \in \mathbb{R}^{|\mathcal{N}|\times d_\text{v}}$ stores the node-specific features, where the $i$th row of the matrix is a $d_\text{v}$-dimensional vector representing the feature vector associated with node $i$.
Likewise the link specific feature vectors are stored in matrix $\bm{E}\in \mathbb{R}^{|\mathcal{E}|\times d_\text{e}}$.

\paragraph{Input layer}
Each initial node embedding vector is derived from the 2-D geometric node coordinates.
Then it is transformed into a $d_\text{hid}$-dimensional vector through a linear layer whose parameters are shared for all nodes.
The initial embedding vector for node $i$ is
\begin{equation}
\xx_i^{0} = \text{LIN}_1(\text{pos}_i).
\end{equation}

To obtain the initial embedding vector for a link, we define an indicator function $I(u,v)$ for link $(u,v)$. 
$I(u,v)=1$ if either node u or v itself is or directly connected to a source or destination node of any commodity flow;
$I(u,v)=0$ otherwise.
The $d_{\text{hid}}$-dimensional embedding vector of link $(u,v)$ is the concatenation of the end point's coordinates, the capacity and the indicator values after linear transforms:
\begin{align}
    \yy_{(u, v)}^{0} = & \;
    \text{LIN}_1(\text{pos}_u)\ ||\ 
    \text{LIN}_1(\text{pos}_v)  \nonumber \\
    & \ ||\ \text{LIN}_2(c(u, v))\ ||\ 
    \text{LIN}_3(I(u, v)).
\end{align}

\paragraph{Graph convolution}
The node and link embedding vectors are updated iteratively for a fixed number of rounds $L$.

In the $(l+1)$-th iteration, the update is a weighted combination of the neighbors' embedding vectors:

\begin{align}
 \yy_{(u,v)}^{(l+1)} &= f_1^{(l)} (\yy_{(u,v)}^{(l)}, \AGG(\mathcal{M}_{(u,v)}), \xx_u^{(l)}, \xx_v^{(l)}) \\
\xx_{u}^{(l+1)} &= f_2^{(l)}(\xx_u^{(l)}, \frac{1}{\lvert \mathcal{N}_u \rvert}\sum_{v\in \mathcal{N}(u)} \yy_{(u,v)}^{(l)}) \\
\mathbf{z}_{(u,v),(u\prime, v\prime)} &= \MLP( (\xx_u \mathbf{W}_6 + \xx_{v\prime} \mathbf{W}_7) || (\mathbf{x}_v \mathbf{W}_8 + \mathbf{x}_{u\prime}\mathbf{W}_9))
\end{align}
where
\begin{align}
f_1^{(l)}(\mathbf{a}, \mathbf{b}, \mathbf{c}, \mathbf{d}) = 
\MLP (\mathbf{a} +  \MLP((&\mathbf{a} \mathbf{W}_1^{(l)} + \mathbf{b} \mathbf{W}_2^{(l)}) \nonumber \\
&\Vert (\mathbf{c} \mathbf{W}_3^{(l)} + \mathbf{d}\mathbf{W}_4^{(l)})),    
\end{align}
\begin{align}
 f_2^{(l)}(\mathbf{a}, \mathbf{b}) &= \MLP (\mathbf{a} +  \MLP(\mathbf{a} + \mathbf{b} \mathbf{W}^{(l)}_5)),
\end{align}
and
\begin{align}
\AGG(\mathcal{M}_{(u,v)}) &= \MLP\left(\sum_{(u\prime,v\prime) \in \mathcal{M}_{(u,v)}}\mathbf{z}_{(u,v),(u\prime, v\prime)} \circ \yy_{(u\prime, v\prime)}\right).  
\end{align}
$\mathcal{M}_{(u,v)}$ is the set of neighborhood links: the links with nodes that are within a distance $L_{M}$ of $u$ or $v$, and 
$\bm{W}^{(l)}_{1-9}$ for all $l$ are trainable parameters associated with each layer.

\begin{figure}
	\centering
	\includegraphics[height=4.5cm]{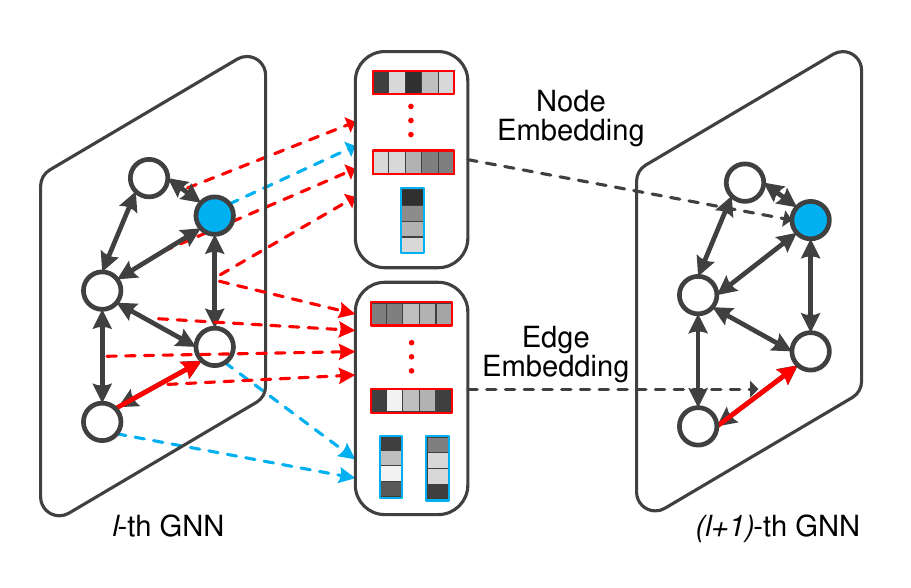}
	\caption{The proposed graph convolution layer operation.}\label{fig:gnn}
\end{figure}

The above update takes the current embedding vectors from the neighborhood of each node or edge, linearly transforms it, and adds the result to the embedding vector of the current node or edge.
It takes a form which is similar to that in \cite{ying2018graph}, but is important in the following ways.
First, the graph convolution is \emph{link centric}.
As our problem is focused on link-level decisions, and nodes do not provide much information other than the specify the flow endpoints, much of the information is around the links and their neighbors instead of nodes;
second, to differentiate the importance of neighbors, each neighbor now has an individual weight factor;
additionally, neighbors of further hops are considered while the most common application only interacts with one-hop neighbors.

\paragraph{Embedding data demands}
To encode the information of demand sets $\mathcal{C}$, we treat a node pair as if it is a virtual link.
we use a feed-forward network with set-invariant properties~\cite{zaheer_deep_2017} using the embedded nodes as the input with
$ \bm{q}_{\mathcal{C}} = \sum _{k \in  \mathcal{C}} \tau_{q} (\xx^{(L)}_{s_k} \Vert \xx^{(L)}_{t_k}),$ where $\tau_{q}$ is a learnable MLP block.
This operation concatenates the final node embedding vectors of source and destination nodes.
The final results are added as a vector representation of the demand.

\paragraph{Link prediction with attention}
The prediction $\hat{\bm{y}}\in \mathbb{R}^{|\mathcal{E}|}$, with elements in $[0,1]$ indicating the likelihood that the link is going to be relevant in the master problem, is generated with the attention mechanism. 
Specifically, 
\begin{equation}
 \hat{\bm{y}} = \text{ReLU}(
  \text{Attention}(\bm{q}_{\mathcal{C}}\bm{W}_{q}, \bm{Y}^{(L)}\bm{W}_{v}))   
\end{equation}
with $\text{Attention}(\mathbf{x}, \bm{Y}) \triangleq \mathbf{x}^T\mathbf{Y},$
where the demand set vector $\bm{q}_{\mathcal{C}}$ and the edge embedding are first converted to an equal dimension by the coefficients $\bm{W}_{{q}}$ and
$\bm{W}_{v}$, and then use inner product followed by the non-linear ReLU activation function to get the final prediction.

Intuitively, this operation measures for each link how relevant they are under the current network topology and the given demand set, where the score is given by the inner product. In our implementation, links with a score higher or equal to a threshold $\alpha$ will be maintained; otherwise, the links are pruned from the topology. Note that $\alpha$ is a parameter that will impact the tradeoff between reduced network size and the quality of the final optimization solution, which will be tuned during experiments. 

\begin{figure}
    \centering
    \includegraphics[width=0.9\linewidth]{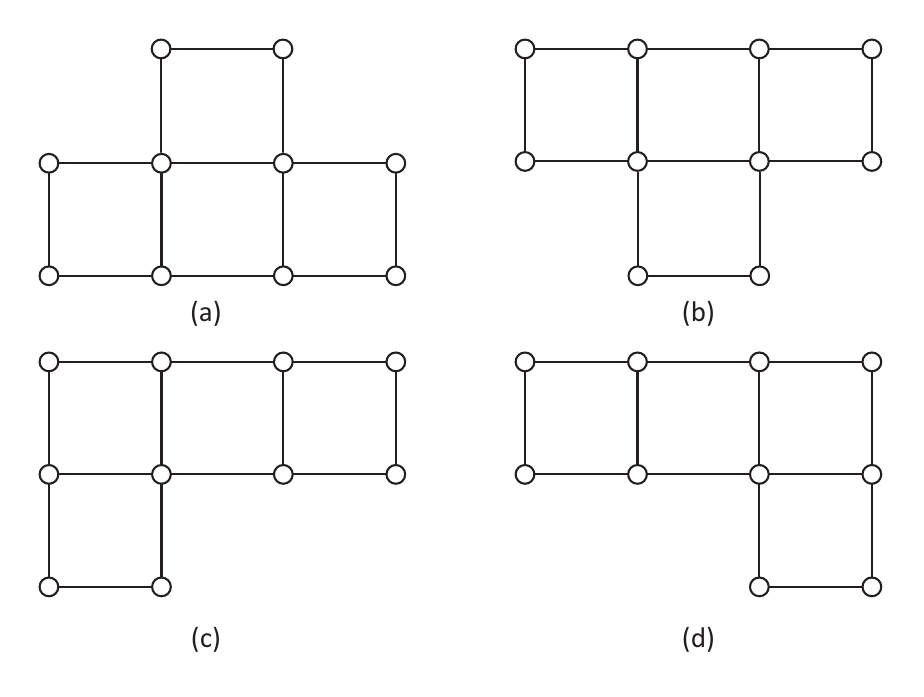}
    \caption{Four types of topology}\label{fig:embed_topo}
\end{figure}

\begin{figure}
	\centering
	\includegraphics[width=0.9\linewidth]{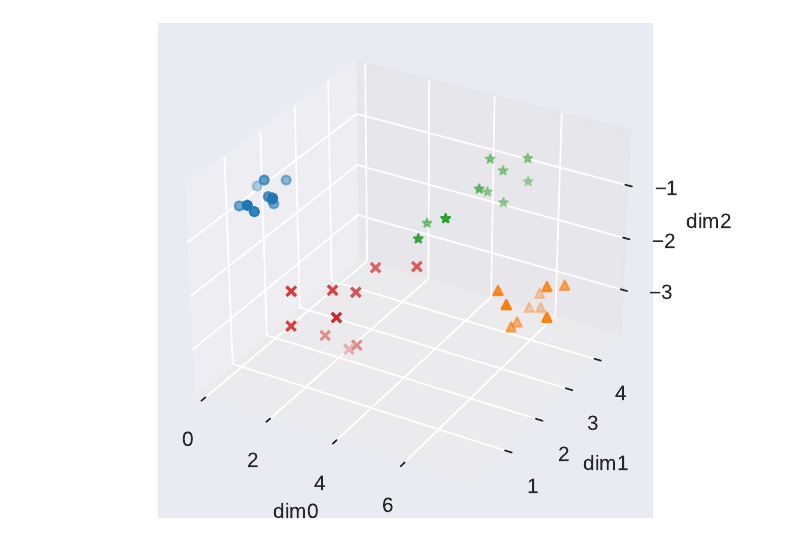}
	\caption{Illustration of the capability of embedding vector in encoding and differentiating the settings of network topologies and link capacities.}\label{fig:embedding}
\end{figure}

As an example, we give a visual demonstration of how the topology-aware embedding distinguish and recognize network instances by illustration.
We generate $4$ network instances, as shown in \cref{fig:embed_topo}  with different topology, and for each we randomly perturb its link capacity for 10 times.
For each of the 40 instances, we apply the topology-aware graph embedding processing and take the sum of the embedding link vectors, which is projected to a 3D vector.
In \cref{fig:embedding}, we plot the projected vectors and can see that they form neighborhoods and instances with closer distance in the projected space correspond to instances that are close in topology and capacity value.

\subsection{Customized learning techniques}
As the problem we study is of huge scale and of irregular form, we take these additional steps to ensure efficient training.

\paragraph{Curriculum training}
We take a \emph{curriculum training} \cite{bengioCurriculumLearning2009} approach to organize the samples and conduct training. The training samples obtained over different network sizes with a different number of commodity demands can be interpreted as have different levels of difficulty. A smaller network size with a smaller number of commodity demands has a lower level of difficulty. Given a setting with a certain network size and number of commodity demands, we will generate training instances by properly arranging the locations of commodity source and destination node so that the training samples can cover a plenty variety of traffic patterns showing how traffic flow needs to be distributed within the network under different demand and interference scenarios. Furthermore, for cases with the same network size and the same number of commodity demands, the difficulty level is differentiated by the number of independent sets incurred in the optimization solution, which implicitly reflects the interference relationships in the case being evaluated.
With all these efforts, we then utilize samples of different difficulty for different training epochs.
The difficulty increases as the training goes further.

\paragraph{Loss function}
Since the ultimate test of learning performance, which is the optimum solution to the problem (\ref{op:formulation}) given the current estimate of link usefulness, is a non-differentiable function of the model parameters, we instead use a differentiable proxy measure, a modified cross-entropy.
It is the sum of binary cross-entropy of the individual link's usage distribution as
\begin{equation}
\mathcal{L} = \sum_{i=1}^{|\mathcal{E}|} w_{i}\big(  y_i \log \hat{y_i} + (1 - y_i) \log (1 - \hat{y_i})  \big),    
\label{lossfunc}
\end{equation}
where $y$ and $\hat{y}$ are respectively the actual and predicted link values; $w_{i}$ is the link-wise weight coefficient. This value is calculated on batch sample basis when using SGD (stochastic gradient descent) algorithm to minimize.

Note that the purpose of (\ref{lossfunc}) is essentially to facilitate performing a multi-label classification on the links: given a network topology $G$ and a demand set $\mathcal{C}$, for each of the link $l$, the goal is to use neural networks to approximate the conditional probability $\hat{y}_{l} = p_{l}(\text{use} = 1 | G, \mathcal{C})$.
The likelihood of observing the link usage pattern is therefore
$\prod_{l} \hat{y}_{l}^{{y_{l}}} (1 - \hat{y}_{l})^{1 - y_{l}}$.
Equivalently, its negative logarithm is the exact form of cross-entropy.

The motivation of incorporating the link-wise weight coefficient is to mitigate the prediction errors over those highly important links. Specifically, we assign the weight coefficient over link $l$ as 
\begin{align}
    w(l) & = \beta \times \text{normalized flow}  \\
         & = \frac{\beta\sum_{k=1}^K f_k(l)}{\sum_{k=1}^K r_k}.
\end{align}
That is, the link weight coefficient for link $l$ is proportional to the ratio between the total network flow carried over link $l$ and the total commodity throughput over the network. The parameter $\beta$ can be considered as a parameter that can tune the sensitivity of the weight coefficient in affecting the learning performance. $\beta$ can be adjusted in practice; in our experiment we find setting $\beta=10$ is a good choice.

\paragraph{Feasibility guarantee}
Note that the trained machine predicts important links in the probabilistic sense. 
There is a possibility that inaccurately pruned links might hurt the network connectivity and thus impact the feasibility of the network optimization. In our implementation, given a topology, we pre-compute a path for each commodity flow. 
The subset of links finally selected (which will be fed to the optimization solver) is the union of the subset of links survived from machine pruning and the links from all pre-computed commodity shortest paths. 
With such an approach, the union of pre-computed shortest paths ensures network connectivity, while the incorporation of important links predicted by machine ensures the quality of optimization. We would like to emphasize that the above operations for feasibility guarantee will not cause any extra cost to the training procedure. The pre-computation of commodity paths can be interpreted as a pre-processing step when the trained machine is applied over a new topology. If there are a large number of application scenarios over the same topology to be evaluated, this pre-processing is just a one-time cost.

\section{Numerical Experiments \label{secNum}}

\subsection{Experiment setup}\label{expset}

To obtain training data in a topology-aware context, we compute a large number of optimization instances over various network topologies and commodity flow deployments by the DCG algorithm \cite{cheng2014systematic}. 
All these instance solutions will be mixed together to form a training data set to conduct supervised learning. 
The trained machine will then be applied for link prediction over different new topologies and commodity flow cases to examine the topology-aware efficiency and robustness of our learning methods.

\paragraph{Datasets}
The network instances are generated following different rules:
In \texttt{random} dataset, the network is a \emph{random geometric} graph:
a given number of nodes are uniformly randomly placed within a square area, with minimum and maximum distances;
In the \texttt{grid} dataset, the nodes are placed on a rectangular grid with a unit Gaussian random perturbation to model the error in deployment.
The nodes are \emph{connected} to as many neighbors as possible within their transmission range, in the sense that any node within the transmission range can be chosen as the next hop by the scheduling process.
For each case, the nodes with data demands are selected randomly, and the total number of demands range from $1-5$.
We generate data samples with $10$, $30$ and $50$ nodes to represent the small, medium and large networks scales. 

Note that the separate instance types are for the purpose of ensuring there are data belonging to the two commonly encountered scenarios;
in the training process there is no distinction in how the model calculates the decisions based on the instance generation rules.
We list the parameters used for instance generation in the \Cref{tab:para-gen}.
\begin{table}[h]
\begin{tabular}{@{}lr@{}}
\toprule
Name               & Value \\ \midrule
Transmission Range & 70 m \\
Interference Range & 50 m \\
Minimum Distance   & 20 m \\
Area Side Length  & 1000 m \\
Transmission Power  & 1mW \\
Number of Nodes    & 10,30,50 \\
Number of Flow Commodities & 1-5 \\
\bottomrule
\end{tabular}
\caption{The list of parameters used in instance generation and training.}\label{tab:para-gen}
\end{table}

\paragraph{Comparison}
We experiment with three additional methods to compare with our TADL method:
\begin{itemize}
  \item \texttt{SAGECONV}, the embedding units in TADL are replaced with the node-based graph invariant embedding model developed in~\cite{ying2018graph};
  \item \texttt{DL-ADJ}, the learning framework in  \cite{ liu_deep_2018} is enhanced with adjacency matrix as topology information;
  \item \texttt{T-Blind}, a topology-blind scheme where some links are arbitrarily activated (without machine learning) in addition to the pre-computed shortest paths that ensuring feasibility.
\end{itemize}
In comparison, the four methods involved are evaluated with the same training and verification datasets.

In the experiments, we examine the performance of our proposed mainly by two metrics.
Essentially we aim to quantify how much speed up the neural model can achieve by cutting irrelevant network links, at the loss of what fraction of the optimum value.
These metrics are defined as follows.

\paragraph{Approximation ratio}
It is defined as the ratio between the optimal network flow computed over the reduced problem instance (denoted as $\text{OPT}_{\text{pruned}}$) and that computed over the original problem (denoted as $\text{OPT}_{\text{original}}$), that is,
$r_{\text{approx}} = \text{OPT}_{\text{pruned}} / \text{OPT}_{\text{original}}.$
It shows how the pruned problem instances approach the optimal network capacity.

\paragraph{Computation time reduction}
Given an optimization instance, we use $t_{\text{org}}$ to denote the computation time solving the original problem. When a pruning method as described in Subsection~\ref{expset} is implemented, the effective computation time will be
$t_{\text{ML}}= t_{\text{setup}} + t_{\text{inference}} + t_{\text{reduced instance}},
$ where
$t_{\text{setup}}$ is the time to compute the commodity shortest paths for feasibility guarantee;
$t_{\text{inference}}$ is the time for the trained machine to prune links (which is $0$ for the T-Blind algorithm);
$t_{\text{reduced instance}}$ is the time for solving the reduced-sized instance. The computation time reduction radio is defined as $r_{\text{red}}=(t_{\text{org}}-t_{\text{ML}})/t_{\text{org}}$.

\begin{table*}[ht]
\small
\centering
\begin{subtable}{\linewidth}
    \centering
\begin{tabular}{@{}lccccccccc@{}}
\toprule
 & \multicolumn{3}{c}{10 nodes} & \multicolumn{3}{c}{30 nodes} & \multicolumn{3}{c}{50 nodes} \\
\cmidrule(l){2-4}\cmidrule(l){5-7}\cmidrule(l){8-10}
No. of commodities & 1         & 3         & 5        & 1          & 3         & 5         & 1          & 3        & 5   \\
         \midrule
   TADL     &  0.96   & 0.92  & 0.84 & 0.96 & 0.82 & 0.72 & 0.98 & 0.87 & 0.79 \\
   SAGECONV &  0.82   & 0.87  & 0.81 & 0.84 & 0.77 & 0.72 & 0.74 & 0.73 & 0.71   \\
   DL-ADJ   &  0.78   & 0.71  & 0.70 & 0.71 & 0.67 & 0.62 & 0.72 & 0.69 & 0.58   \\ 
   T-BLIND  &  0.67   & 0.72  & 0.72 & 0.68 & 0.64 & 0.66 & 0.56 & 0.51 & 0.43   \\  
\bottomrule
\end{tabular}
\caption{Approximation Ratio}\label{tab:snapshot-ar}
\end{subtable}

\vspace{0.3cm}

\begin{subtable}{\linewidth}
    \centering
    \begin{tabular}{@{}llccccccccc@{}}
    \toprule
        & \multicolumn{3}{c}{10 nodes} & \multicolumn{3}{c}{30 nodes} & \multicolumn{3}{c}{50 nodes} \\
        \cmidrule(l){2-4}\cmidrule(l){5-7}\cmidrule(l){8-10}
        No. of commodities  & 1         & 3         & 5        & 1          & 3         & 5         & 1          & 3        & 5   \\
        \midrule
TADL        & 0.61 & 0.47 & 0.28  & 0.45  & 0.4   & 0.17 & 0.34 & 0.27 & 0.18 \\
SAGECONV    & 0.44 & 0.39 & 0.16  & 0.4   & 0.33  & 0.16 & 0.28 & 0.15 & 0.13\\
DL-ADJ      & 0.38 & 0.32 & 0.18  & 0.23  & 0.11  & 0.12 & 0.22 & 0.27 & 0.12\\
T-BLIND     & 0.42 & 0.40 & 0.37  & 0.48  & 0.35  & 0.35 & 0.37 & 0.33 & 0.04\\
\bottomrule
\end{tabular}
\caption{Time Reduction}\label{tab:snapshot-tr}
\end{subtable}
\caption{Approximation ratio and time reduction ratio comparison for randomly generated cases with different instance sizes. Both performance metrics are the larger the better.}\label{tab:acc}
\end{table*}

\subsection{Time cost reduction and optimality in individually trained scenarios}
To give a general idea of how the model performs, in this experiment, 
we train and test models on different instance sizes. 
For each size, instances have different flow demands, node and link numbers to observe the performance under different data configurations.
The testing is performed with the samples that are reserved and not seen by the training process from the same dataset.

In \Cref{tab:acc}, we demonstrate typical values of approximation ratios and time reduction in randomly generated networking cases.
The data is sub-grouped according to the number of the network nodes number of commodity flows.
From the results, we can tell that our TADL methods robustly achieve a good tradeoff between optimization quality and complexity reduction.
The two methods with embedding (i.e., TADL and SAGECONV) always perform better than the other two methods. 
In almost all the scenarios, the adjacency matrix method DL-ADJ perform similar or even worse than T-Blind. 
The results confirm our analysis before that the change of an adjacency matrix cannot give an accurate indication of the topology change and thus hard to extract meaningful topology related information to facilitate link prediction. 
Note that all four methods in our experiments maintain a usable approximation ratio due to our feasibility guarantee mechanism.

It can be also seen that generally as the number of flow requirements increases, the needed solution time increases while the problem value decreases.
This is because with a higher number of flow demands, the overall scheduling is more fragmented as at any time only a link can only carry one flow.
This causes the scheduling task to take more iterations to discover a good set of ISs, and that there are more links that need to be considered in the solution.
Accordingly, the time reduction and approximation ratio in low flow count cases are the most significant because the model can accurately infer the needed links.
When there are a relatively large number of commodity flows in the network, it can be expected that a big portion of the links need to be used, and thus the margin for effective pruning will be smaller and the chance of inaccurate pruning will be increased. 
We indeed observe the slight decrease of $r_{\text{approx}}$ and $r_{\text{red}}$.

\subsection{A practical office setting}
\begin{figure*}
         \centering
         \includegraphics[height=6.5cm]{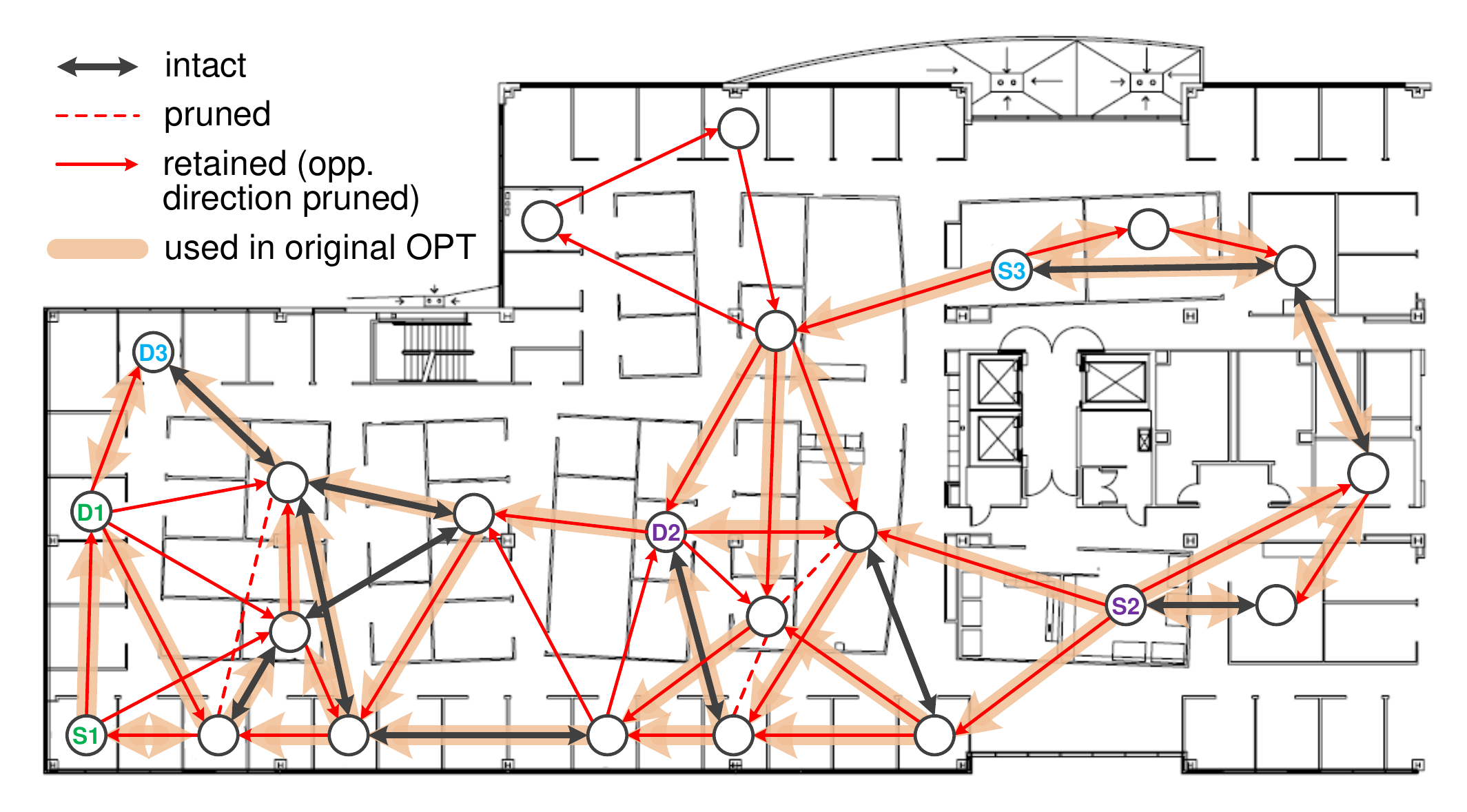}
         \caption{A wireless mesh topology in an office setting constructed in \cite{draves_routing_2004}.TADL link prediction and the optimal link set are shown.}
         \label{fig:office}
     \end{figure*}

To further examine the robustness of TADL, we apply it over a wireless mesh  topology with 3 commodity flows, in a practical office setting which was studied in \cite{draves_routing_2004} as shown in Fig.~\ref{fig:office}. 
In the topology, each edge represents a bi-directional link, thus giving 96 unidirectional links in total. Here, we didn't do any retraining; we directly use the machine trained over a 25-node instance dataset to process the office topology. 
TADL generates a reduced problem of 56 links with the approximation ratio $r_{\text{approx}}=0.97$ and the time reduction ratio $r_{\text{red}}=0.42$. Fig.~\ref{fig:office} also indicates the exact set of links that are activated in the optimal solution from the original problem to benchmark the prediction accuracy. We can tell that TADL only incorrectly prunes a few links and include a small set of redundant links. 

\subsection{Training Convergence and loss function comparison}
\begin{figure}
    \centering
    \includegraphics[trim={1cm 0 0 1cm},clip, width=0.9\linewidth]{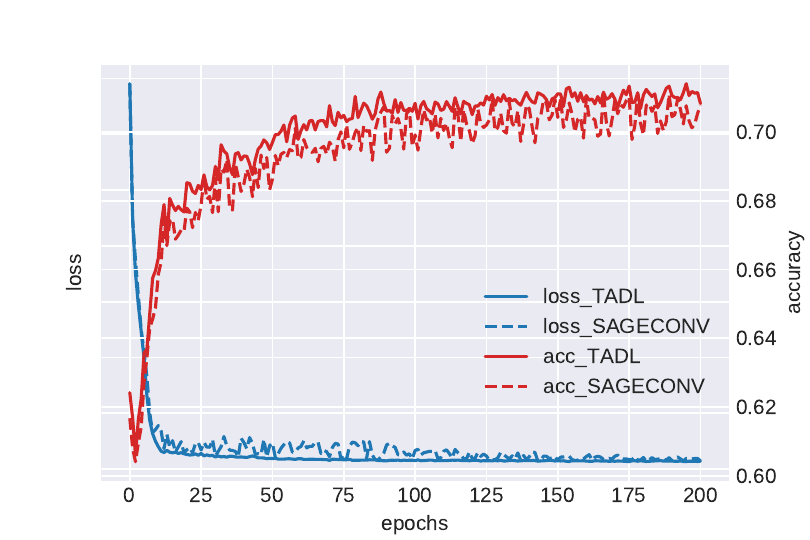}
    \caption{The loss and accuracy metric during training.}\label{fig:loss}
\end{figure}

The training loss and the prediction accuracy of the proposed method and the SAGECONV are illustrated in~\cref{fig:loss}.
The data was taken when training the model on 30-node datasets with a mixed number of demands, which is generally more difficult for the model to differentiate and also its size is commonly encountered in the application.
We can observe that for our method, the descent of loss and the ascent of prediction accuracy happens at a higher slope and generally a few epochs earlier than the other studied method, despite using similarly tuned learning rate and batch size settings. These indicate that our proposed method is more suited with this problem data.

\begin{figure}
\centering
\includegraphics{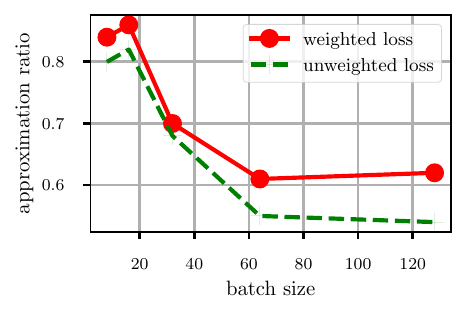}
\caption{
  Effects of batch size and loss choice.
  We observe that the batch size demonstrates a U-shaped curve, indicating that between 8-32 there exists a sweet spot.
  Weighted losses that consider the sample weights also improves the learning results.
}\label{fig:eigenval-batchsize}
\end{figure}

In addition, choosing a \emph{sample weighted} loss function is clearly observed as a better choice from \cref{fig:eigenval-batchsize}.
Under different batch size tuning, there is a performance gap of around 3\% between the weighted and unweighted versions.
This can be explained by the fact that in this specific application the class imbalance in the data is serious enough that additional weighing on the links to be used can help contribute to the final metric which the user cares the most.

\subsection{Model Performance with Sample Count}
Since learning-based prediction is inherently a data-oriented approach, the model performance generally improves as the number of training samples increases.
To verify this, in this experiment, several models with the same setting are trained with different numbers of samples from the same data distribution.
After training, the performance of the models are tested on the same set of non-training cases' data.

\begin{figure}
  \centering
  \begin{subfigure}{\linewidth}
    \centering
    \includegraphics[width=0.8\linewidth]{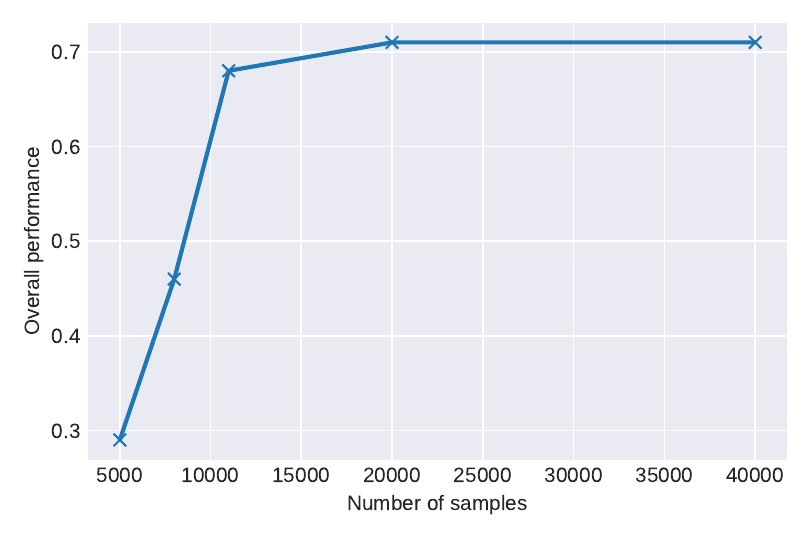}
    \caption{10 node networks.}
  \end{subfigure}
  \begin{subfigure}{\linewidth}
    \centering
    \includegraphics[width=0.8\linewidth]{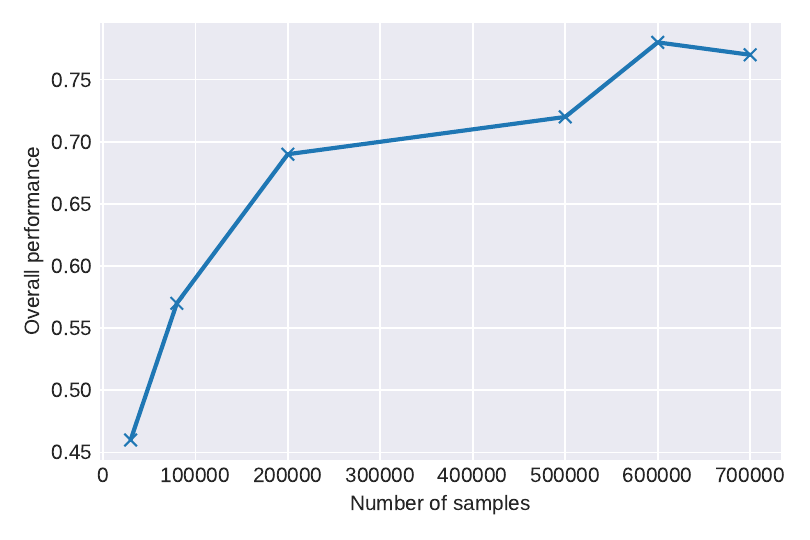}
    \caption{30-node networks}
  \end{subfigure}
  \begin{subfigure}{\linewidth}
    \centering
    \includegraphics[width=0.8\linewidth]{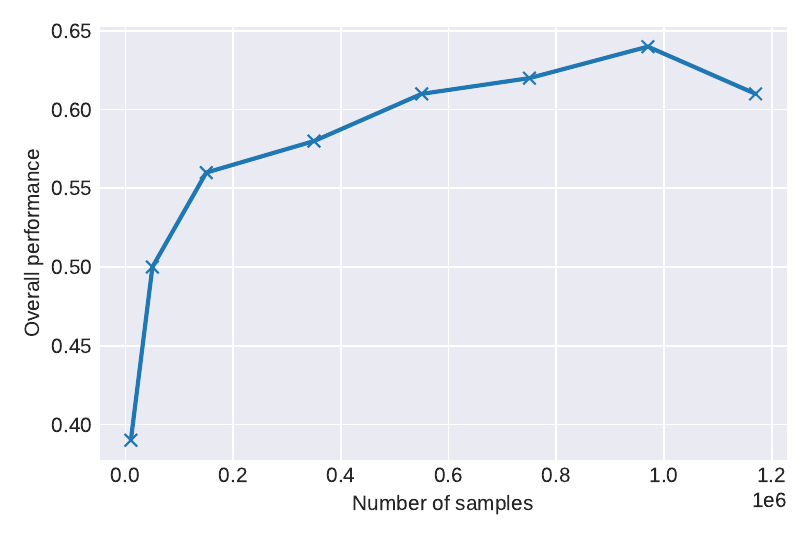}
    \caption{50-node networks}
  \end{subfigure}
  \caption{Model performance with changing number of training samples. The overall performance is given as the product of approximation ratio and time reduction, and the larger the better. This figure cannot be greater than 1.}\label{fig:perf-samplecount}
\end{figure}

In \Cref{fig:perf-samplecount}, we consider the product of approximation ratio and the computation time reduction as a metric of the learning quality, and plot this figure of merit against the number of training samples used.
We repeat the same process for instances of different sizes.

The results confirm that the performance typically saturates at a level that goes higher as the number of samples increases. 
This shows that the model's capacity is not fully utilized when the number of training samples is not sufficient, and that before a certain point, more data is indeed better.
But afterwards, the performance metric starts to either oscillate or even increase.
This can be attributed to the model becoming unable to make useful correlations just based on the training samples provided to achieve further improvement on the testing data.
More data beyond this point is not useful unless the model's parameters are changed.

\section{Related Work}
\label{related}

\textbf{Conventional wireless network optimization}
Wireless network optimization had been a key research area in the recent two decades. The basic methodology is to compute the resource allocation aspects such as channel assignment, base station association, scheduling and power control using various mathematical programming algorithms~\cite{han2008resource}. Due to the complex inference relationship, wireless network optimization is NP-hard in general, and the major thread of efforts in the community is the development of various approximation algorithms~\cite{jainImpactInterferenceMultiHop2005, georgiadis2006resource}. The studies had also been extended from single-radio single-channel context to complex multi-radio multi-channel context~\cite{kumar_algorithmic_2005,draves_routing_2004, liu_joint_2017}. 
A particular issue inspiring the machine learning study in \cite{ liu_deep_2018} and this paper is that a new optimization problem instance is always solved either from scratch or with a trivial re-optimization approach~\cite{bertsekas1998network}; machine learning aims to exploit the historical computation effort to benefit new optimization instances.

\textbf{Data-driven solutions to network problems}
Aspects of network design problems can be cast as optimum control problems, and there have been attempts to apply machine learning methods as a way to discover heuristic algorithms from data.
\cite{liu_deep_2018} typifies the supervised approach to optimize the flow scheduling in wireless ad-hoc networks.
Another approach as exemplified in the use of actor-critic style models is reported in  \cite{chen2018auto}, where the neural model optimizes the data flow path in a data center network.
Similar ideas appear in a series of recent online network control problems \cite{mao2016resource, jiang2017pytheas, chen2018auto}.

\textbf{Neural network solutions in solving combinatorial problems}
Recently deep learning-based methods are attempted to solve combinatorial optimization problems.
By using sequential modeling and graph neural networks, end-to-end solutions can be obtained.
Pointer network \cite{vinyalsPointerNetworks2015}, a model based on attention multi-head attention mechanism, solves variable-sized combinatorial problems by using the attention scores as selection criteria, and this approach is shown to achieve reasonable performance level with classic problems including Traveling Salesman Problem and Delaunay triangulation.
This idea is further expanded to solve a vehicle routing problem \cite{koolAttentionLearnSolve2019}.
Using the problem graph instance as an input to a transformer, the output is determined by the embedded node vectors.
On the outer level, the model is further trained by a policy gradient reinforcement learning algorithm REINFORCE.

Another line seeks a proper representation of the graphical structure in the neural network context. 
It is shown that by using a structure-aware model \texttt{structure2vec} \cite{daiDiscriminativeEmbeddingsLatent2016, daiLearningCombinatorialOptimization2017}, the neural networks can learn to build up approximate solutions by iteratively adding new nodes to an existing partial solution, with the necessary problem-specific helper functions.

\section{Conclusions}\label{summary}

This work contributes a topology-aware approach to train a DL machine which can robustly predict important links (and thus prune non-critical links) to facilitate wireless network optimization over dynamic network topologies. Efficient embedding techniques are developed to address the fundamental issue of index-independent topology representation. Our method can work in a complementary manner with the traditional theme of approximation algorithms, to further reduce computation complexity from a new dimension by leveraging historical computation data. As a next step, we will conduct a further in-depth topology-aware study in the more generic setting of multi-radio multi-channel wireless networks.

\printbibliography%

\end{document}